\documentclass[aps,prl,twocolumn,groupedaddress,showpacs,amsfonts,floatfix]{revtex4}
\usepackage{graphicx}
\begin{document}
\title{Critical phenomena of thick branes in warped spacetimes}
\author{Antonio Campos}
\affiliation{Institute of Cosmology and Gravitation,
             University of Portsmouth,
             Portsmouth PO1 2EG,
             United Kingdom}
\date{\today}
%\begin{abstract}
%We have investigated the effects of a generic bulk first-order phase 
%transition on the dynamics of thick Minkowski branes in warped 
%geometries.
%We have found that, as occurs in an Euclidean spacetime,
%a film of disordered phase appears between two ordered bulk phases
%as the phase transition is approached.
%Thus, an interface separating two ordered phases split into 
%two interfaces with a disordered phase in between.
%A remarkable and distinctive feature is that the critical temperature
%of the phase transition is lowered due to pure geometrical effects.
%We have studied a variety of critical exponents which describe
%the universal aspects of the transition and the evolution of
%the transverse-traceless sector of the metric fluctuations. 
%\end{abstract}
\begin{abstract}
We have investigated the effects of a generic bulk first-order phase 
transition on thick Minkowski branes in warped geometries.
As occurs in Euclidean space, when the system is brought near the 
phase transition an interface separating two ordered phases splits 
into two interfaces with a disordered phase in between.
A remarkable and distinctive feature is that the 
critical temperature of the phase transition is lowered due to pure 
geometrical effects.
We have studied a variety of critical exponents and the evolution of
the transverse-traceless sector of the metric fluctuations. 
\end{abstract}
\pacs{11.10.Kk,04.50.+h,05.70.Jk,11.27.+d}
\maketitle

The physics of branes in higher dimensional theories has 
been the subject of renewed interest in the past years,
mostly, because they provide a novel approach for 
resolving the cosmological constant and the 
hierarchy problems \cite{ArkDimDva:1998,RanSun:1999a}.
An intriguing feature of this type of scenario is 
the possibility of localizing a massless graviton on the
brane, and then reproducing effectively four-dimensional gravity at 
large distances, due to the warp geometry of the spacetime
\cite{RanSun:1999b}.
Furthermore, these models have also inspired the idea in which 
our universe is the result of the continuous collisions of branes
\cite{Ekpyrotic:2001}.
A fundamental ingredient of this cosmological model is the 
nucleation and splitting of branes.
Then, it is interesting to consider the universal aspects of 
thick branes \cite{DeWFreGubKar:2000} splitting in a 
warped bulk.
In this work we present a field theory description for a feasible
realization of such a mechanism.
Using superpotential techniques
\cite{DeWFreGubKar:2000,Superpotential:1999},
some exact solutions of the equations of motion for the coupled 
scalar field plus gravity system have been studied in 
\cite{TB:2000}. 
Analytical solutions for giving potentials have been described in 
\cite{Exact_sol:1999}.
The probability of tunneling has been considered in 
\cite{Tunneling:2000} and some geometrical aspects 
of brane collisions have been pointed out in 
\cite{Per:2001,collisions:2001}. 

In this letter, we provide a qualitative description of the 
splitting of thick Minkowski branes due to a first-order phase 
transition in a warped geometry. 
For this purpose we have considered a generic potential for a 
complex scalar field that captures the universal 
aspects of the phase transition in the bulk. 
As the system is brought near the critical temperature an
interface interpolating two bulk phases breaks into two
separated interfaces, and a layer of a new phase appears between 
them.
This effect is known in condensed matter physics as
complete wetting \cite{Lip:1982,Lip:1984}.
It also appears at the deconfinement phase transition of 
SU($3$) Yang-Mills theory \cite{FrePat:1989,TraWie:1992} and 
supersymmetric QCD
\cite{SUSY-QCD:1998}.
Our analysis shows the existence of a new effective critical
temperature for complete wetting due to the effective
negative bulk cosmological constant.
This phenomenon reveals a remarkable relationship between 
geometry and the critical behavior of interfaces.
We have been able to characterize the universal aspects
of this critical phenomenon computing the critical exponents
for some relevant quantities.
And finally, we have also studied the behavior of the metric 
fluctuations.
In particular, we have found that the zero mode delocalizes,
high energy modes are scarcely affected, and for low energy modes
the phase changes and the amplitude starts to grow at the moment
of the splitting.
It is interesting to point out that we are going to use a numerical
approach and thereby our solutions do not depend on any
specific superpotential.

We start with a metric of the form 
\begin{equation}
   ds^2
      \ = \ e^{2A(r)}\eta_{ij}dx^idx^j - dr^2,
\end{equation}
where $i=0, 1, 2, 3$, $\eta_{ij}=(1, -1, -1, -1)$, and 
the warp factor $A(r)$ depends only on the coordinate of the 
extra dimension $r$. 
This choice of metric preserves Poincar{\'e} invariance
in the four-dimensional submanifolds of constant $r$.
To build up the thick brane we consider an auxiliary 
complex scalar field $\Phi$ coupled to gravity through
the five-dimensional action
\[
   S
     \ = \ \int d^4x\ dr \sqrt{|g|}
           ( - \frac{1}{4} R
             + \frac{1}{2} g^{\mu\nu} \partial_\mu \Phi^\ast
                                      \partial_\nu \Phi
             - V(\Phi)
             - \Lambda
           ),                                                       
\]
where $R$ is the scalar curvature and $\Lambda$ is a bulk 
parameter that fixes the minimum of the scalar potential $V(\Phi)$.
In flat backgrounds $\Lambda$ is a free parameter that can 
always be used to set the minimun of the potential to zero.
In contrast, this is not always possible if one is concerned to 
find out soliton solutions of the equations of motion
in curved spacetimes.
Since we are particularly interested in the universal aspects of
first-order phase transitions in the bulk we are going
to consider the generic scalar field potential
\begin{equation}
   V(\Phi)
      \ = \  a |\Phi |^2
            -b \phi_R (\phi_R^2 - 3\phi_I^2)
            +c |\Phi |^4,                                           
\label{eq:potential}
\end{equation}
where $\phi_R$ and $\phi_I$ are, respectively, the real and the 
imaginary part of the scalar field $\Phi$. 
This potential has quartic self-interactions and respects 
${\mathbb Z}_3$ symmetry.
The action for a scalar field with a potential of this type 
belongs to the same universality class as the effective action
for theories describing the ${\mathbb Z}_N$ center symmetry of 
a $SU(N)$ gauge field \cite{TraWie:1992}.
So, they can be easily realized if the model supports a bulk
field with a gauge symmetry.
To ensure stability the parameter $c$ must be positive. 
For $a > 0$ the potential has a local minimum at $\Phi^{(1)}=0$
corresponding to a disordered bulk phase.
For $b > 0$ and $a < 9a_c/8$ there are three additional
degenerate global minima corresponding to ordered bulk phases, 
say $\Phi^{(2)}=\phi_o$, 
$\Phi^{(3)}=(-\frac{1}{2} + i\frac{\sqrt{3}}{2})\phi_o$, and
$\Phi^{(4)}=(-\frac{1}{2} - i\frac{\sqrt{3}}{2})\phi_o$, 
where
\begin{equation}
   \phi_o
      \ = \ \frac{3b}{8c}
            \left( 1 + \sqrt{1-\frac{8}{9}\frac{a}{a_c}}\,
            \right),                                                
\label{eq:OP}
\end{equation}
and $a_c = b^2/4c$.
These three minima coexist with the disordered phase when 
$a=a_c$.
Then fixing $b$ and $c$ lead us to interpret $a$ as a bulk 
temperature parameter where $a_c$ is the critical temperature.
The actual variation of $a$ will strongly depend on the 
particular realization of the effective theory. But, since we 
are studying the universal aspects of the phase transition, 
we can treat this parameter as an independent variable.
Note that the effective five-dimensional cosmological constant
is not $\Lambda$ but $\Lambda_{eff} = \Lambda + V(a)$,
where $V(a)$ indicates the value of the scalar potential at 
the global minima, i.e.
$V(\Phi^{(n)})\equiv \phi_o^2 ( a - b \phi_o + c \phi_o^2 )$
with $n$ taking values $2$, $3$ or $4$.

\begin{figure}
\includegraphics*[totalheight=3.3in,width=0.95\columnwidth]{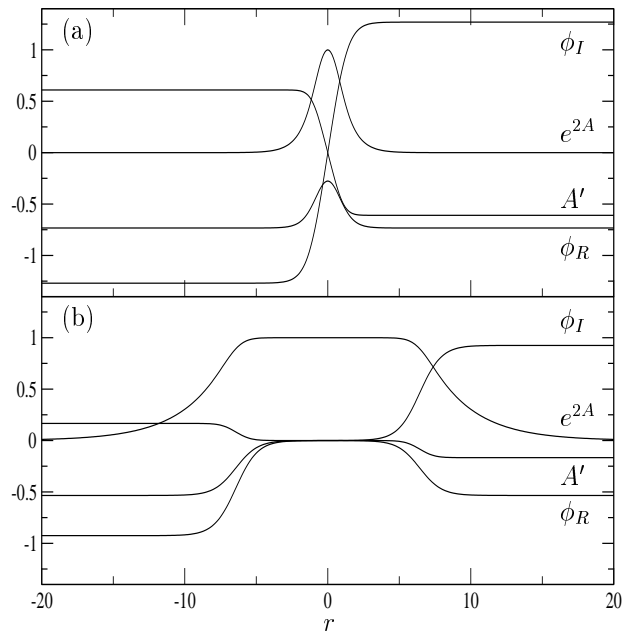}
\caption{\label{fig:profiles}Interface profiles for thick domain 
walls in a warped geometry with a ${\mathbb Z}_3$-symmetric bulk 
first-order phase transition for $b=2$ and $c=1$. 
(a) Away from the phase transition a domain wall interpolates
two distinct ordered phases. 
(b) Close to the critical temperature the wall splits into two 
interfaces with an intervining film of disordered phase in between.}
\end{figure}

The classical equations of motion for planar field configurations 
read as follows
\cite{DeWFreGubKar:2000},  
\begin{eqnarray}
   \phi_R''
      & = & -4 A'\phi_R' + \frac{\partial V}{\partial \phi_R},      
\label{eq:EM_phi_R}
          \\
   \phi_I''
      & = & -4 A'\phi_I' + \frac{\partial V}{\partial \phi_I},      
\label{eq:EM_phi_I}
          \\
   A''
      & = & -\frac{2}{3} ( \phi_R'^2 + \phi_I'^2 ),                 
\label{eq:EM_A}
          \\
   3A'^2
      & = &  \frac{1}{2} ( \phi_R'^2 + \phi_I'^2 )
           - V(\Phi)
           - \Lambda,                                               
\label{eq:CONSTRAINT}
\end{eqnarray}
where the prime denotes differentiation with respect to the
coordinate $r$.
The soliton equations for the flat case are recovered by 
taking the warp factor $A(r)$ to be constant \cite{TraWie:1992} 
(see \cite{SUSY-QCD:1998} for a supersymmetric generalization).
Note also that Eqs.~(\ref{eq:EM_A}) and (\ref{eq:CONSTRAINT}) 
imply that $\Lambda < |V(a)|$ and then the effective five-dimensional 
cosmological constant $\Lambda_{eff}$ is always negative.
We have solved numerically the system of ordinary differential
equations (\ref{eq:EM_phi_R})-(\ref{eq:EM_A}) for 
$\phi_R(r)$, $\phi_I(r)$, and $A(r)$ with boundary conditions 
$\Phi(-\infty)=\Phi^{(4)}$, 
$\Phi(-\infty)=\Phi^{(3)}$,
$A(0)=A'(0)=0$ (see Fig.~\ref{fig:profiles}).
Then, Eq.~(\ref{eq:CONSTRAINT}) has been used in order to 
fix the parameter $\Lambda$ for different values of the parameter 
$a$ as the interface configuration is brought near the phase 
transition.
Because of the ${\mathbb Z}_3$ symmetry there will be 
equivalent solutions for other choices of ordered phases.

Our results reveals that, equivalently to what happens when 
the background spacetime is Euclidean, the ordered-ordered domain
wall splits into two separated ordered-disordered interfaces,
and a layer of disordered phase appears and completely wets the 
two ordered phases as the the phase transition is approached.
Although in both cases this behavior looks quite similar
there is an important distinctive feature. 
For a geometry with a warp factor complete wetting does not occur 
for the critical temperature $a_c$ but for a smaller effective 
temperature $a_\ast$.
As we will see, this effect is closely related to the 
fact that asymptotically the spacetime is Anti-deSitter.  
In order to understand why this effective critical temperature
appears it is necessary to know the behavior of the Euclidean energy
$E(r)\equiv \frac{1}{2}(\phi'^2_R+\phi'^2_I) - V(\Phi)$ as the 
temperature parameter $a$ increases.
It is very easy to see, using the equations of motion 
(\ref{eq:EM_phi_R}) and (\ref{eq:EM_phi_I}), that the derivative of this 
energy along the orthogonal direction to the domain wall is
$E'(r) = -4A'(\phi'^2_R+\phi'^2_I)$.
Then, it has a minimun at $r=0$, $E(0)=\Lambda$, and it
reaches its maximun value at infinity $E(\pm\infty)=|V(a)|$.
In fact, it has the same shape as $3A'^2$ but is displaced 
vertically.
On the other hand, one also has to take into account that the 
stability of a configuration with three coexisting
phases implies that an interface of two ordered phases can
only be completely wet by a disordered phase.
This means that in order to have a disordered phase between
our initial ordered phases $\Phi(0)\rightarrow \Phi^{(1)}$
and $\Phi'(0)\rightarrow 0$ when $a$ increases.
Moreover, $E(0)$ should tend to zero as well as $\Lambda$ 
for complete wetting to occur.
If the background is Euclidean, the energy is a positive 
constant, $E=|V(a)|$, for each value of the temperature 
parameter $a$.
So, since $E$ is zero for $V(a=a_c)=0$, complete wetting 
happens with a critical temperature $a_c$.
However, in the warped case and because $\Lambda$ is always 
smaller than $|V(a)|$, the coexistence of the three phases 
occurs for a critical value $a_\ast < a_c$.
This inmediately translates into the fact that 
the effective bulk cosmological constant asymptotically 
reaches a constant negative value
\begin{equation}
   \Lambda^\ast_{eff}
      \ = \  V(a)|_{a_\ast}
      \ < \ 0.
\end{equation}
Then, one could revert the argument and say that
the existence of a nonzero negative cosmological 
constant is responsable for lowering the critical 
temperature of complete wetting.
One would expect intuitively that the value 
of $a_\ast$ is very close to the critical value $a_c$. 
In this case, it can be easily seen that 
$\Lambda^\ast_{eff}$, $a_c$, and $a_\ast$ are related by,
\begin{equation}
   \Lambda^\ast_{eff}
      \ \sim \  - \frac{a_c}{c} (a_c - a_\ast).
\end{equation}
It is worth noticing that the existence of an effective 
critical temperature prevents the global geometry for becoming
flat as the phase transition is reached. If the contrary would 
have occurred, it would have been a puzzling feature of brane 
models.
In the example plotted in Fig.~\ref{fig:profiles}
the critical temperature is $a_c=1$.
In this case, we have obtained the effective critical temperature 
$a_\ast=0.9223822740$ for $\Lambda < 10^{-7}$ and an effective 
bulk cosmological constant $\Lambda^\ast_{eff}\sim -0.08$.

\begin{figure}
\includegraphics*[totalheight=2.7in,width=0.95\columnwidth]{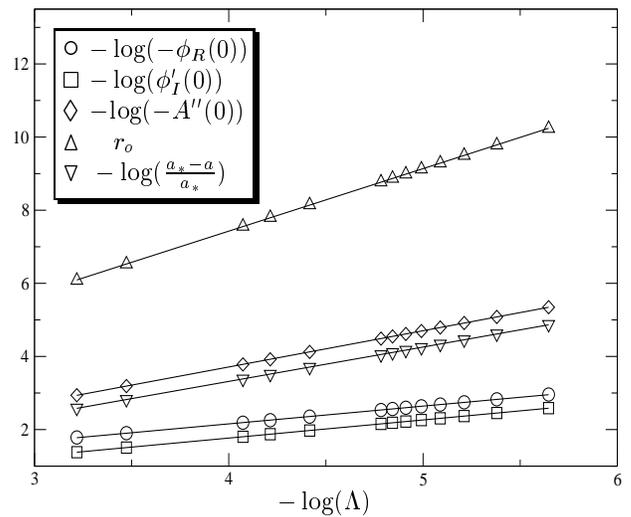}
\caption{\label{fig:crit_exp}Determination of the critical 
exponents for a variety of order parameters of the phase transition.}
\end{figure}

The universal aspects of this critical behavior are 
characterized by several critical exponents.
In Fig.~\ref{fig:crit_exp} we have plotted the numerical results
for the determination of the critical exponents of several
parameters of the phase transition.
As one would expect from the discussion above, the parameter
$\Lambda$ is proportional to the deviation from the effective
critical temperature, $(a_\ast - a)/a_\ast \propto \Lambda$.
On the other hand, the width of the disordered phase layer, 
$r_o$, wetting the two ordered phases diverges logarithmically
$r_o \propto -\log (\Lambda)$.
In addition, for the order parameters at the center of the
wetting layer we obtain $A''(0)\propto\Lambda$, 
$\phi_R(0)\propto\Lambda^{1/2}$, and $\phi'_I(0)\propto\Lambda^{1/2}$.
In condensed matter physics, this type of critical behaviour 
corresponds to systems with short-range interactions \cite{Lip:1984}.  

\begin{figure}
\includegraphics*[height=3.3in,width=0.95\columnwidth]{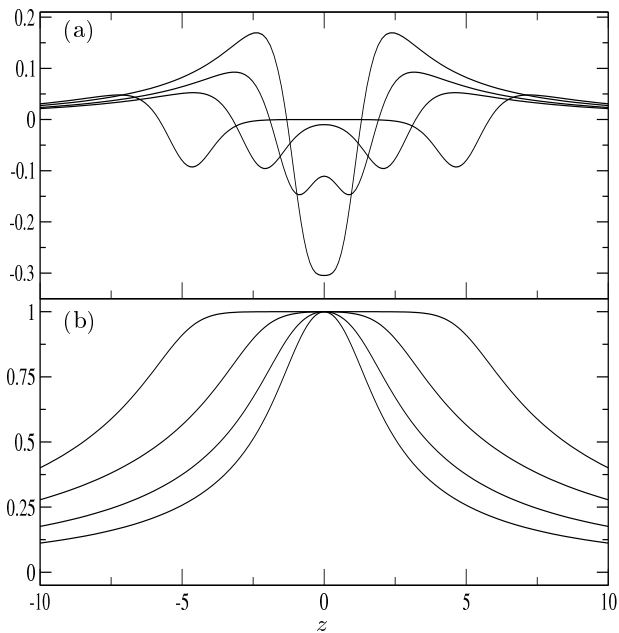}
\caption{\label{fig:volcanos}Shapes for the volcano 
potential (a) and the wave function of the zeromode (b) 
for the transverse-traceless part of small gravitational 
perturbations. 
The critical temperature is approached from bottom to top 
in both plots and we have taken an arbitrary normalization
for the wave function. 
Note that, away from the transition, the zeromode is 
bound to the walls in the fifth extra dimension but, as the
walls peel off and pull apart, it is delocalized.}
\end{figure}

Now, let us consider the metric fluctuations with respect to
this background solution. 
In the axial gauge $h_{i4}=0=h_{44}$ the perturbed metric
can be written as
\begin{equation}
   ds^2
      \ = \ e^{2A(r)}(\eta_{ij}+h_{ij})dx^idx^j - dr^2.
\end{equation}
The transverse and traceless (TT) sector of the perturbations,
$\bar{h}_{ij}$, decouples from the scalar field perturbations
and satisfies the equation \cite{DeWFreGubKar:2000},
\begin{equation}  
   \left( \partial^2_r +4A'\partial_r - e^{-2A}\Box
   \right) \bar{h}_{ij}\ = \ 0.
\end{equation}
This equation can be transformed into a 
Schr$\ddot{{\rm o}}$dinger-like equation by changing to the
coordinate $dz=e^{-A}dr$ and introducing the new variable 
$H_{ij}=e^{-ikx}e^{(3A/2)}\bar{h}_{ij}$,
\begin{equation}
   \partial^2_z H_{ij} 
      \ = \ \left( U(z)-k^2 
            \right) H_{ij}.
\end{equation}
Here, the potential is
\begin{equation}
   U(z)
      \ = \   \frac{9}{4}\dot{A}^2 
            + \frac{3}{2}\ddot{A}
      \ = \   \frac{3}{4} e^{2A}
              \left( 2A'' + 5 A'^2
              \right),
\end{equation}
where the dot now stands for the derivative with respect to the 
new coordinate $z$. 
The zero mode $k=0$ has the analytical solution 
$H_{ij}=N_{ij} e^{(3A/2)}$, with $N_{ij}$ a 
normalization factor.
As shown in Fig.~\ref{fig:volcanos}, away from the phase 
transition the potential in the Schr$\ddot{{\rm o}}$dinger
equation for the TT sector of the metric perturbations has 
the shape of a volcano potential.
However, after the wall splits off, the shape becomes that 
of a double-well-type potential with a depth decreasing
as the two walls separate.   
One can observe that the critical behavior delocalizes
the zero mode, but it does not introduce
any gap between the zero mode and the continuum
spectrum. 
Furthermore, the zero mode is the lowest energy eigenstate
and then there are no instabilities coming from tachyonic
modes.

\begin{figure}
\includegraphics*[height=3.3in,width=0.95\columnwidth]{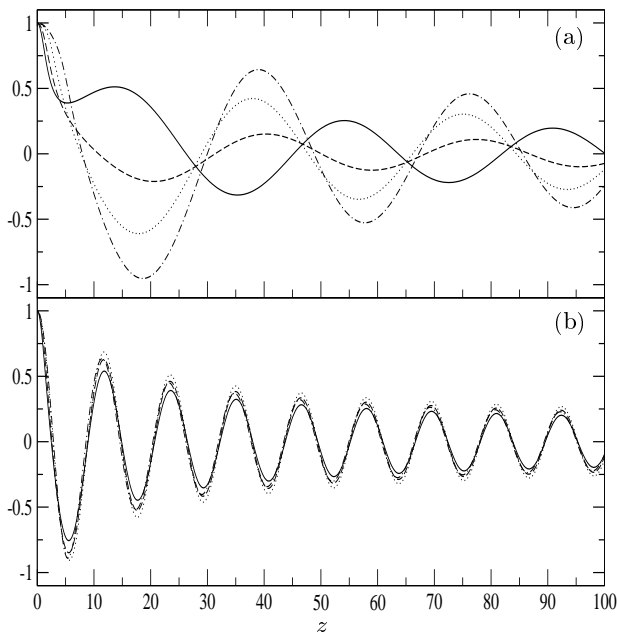}
\caption{\label{fig:non-zero}Wave functions for nonzero modes 
of metric fluctuations with an arbitrary normalization. 
(a) Modes with energy smaller than the maximun of the 
volcano potential, and (b) modes with larger energy. 
The plots represent different stages of the phase transition
in the following order: solid line, dashed line, dot line, and 
dashed-dot line. Being the latest the closest to the critical 
temperature.}
\end{figure}

In Fig.~\ref{fig:non-zero}, we have also plotted the wave functions
of two different types of nonzero modes.
In Fig.~\ref{fig:non-zero}(b), we have considered the case of modes 
with energy larger than the maximun of the volcano potential, and 
in Fig.~\ref{fig:non-zero}(a), those with a smaller energy.
The first thing to note is that the wave functions get broader
as the mode has lower energy.
Modes with larger energy barely get affected by the splitting
of the interfaces.  
On the contrary, the phases of the waves for modes with a small 
energy change at the moment of the nucleation, increasing their
amplitude afterwards.

In conclusion, we have investigated the critical behavior of
coexisting interfaces in an asymptotically anti-de Sitter bulk. 
Because a film of disordered phase is formed between two ordered
phases, this mechanism of complete wetting gives a generic field 
theory description of the splitting of a thick domain wall, 
which is a fundamental ingredient for the brane cosmological 
models discussed in \cite{Ekpyrotic:2001}.
Moreover, we have found that naturally an effective critical 
temperature shows up due purely to the global geometry.
This is a remarkable result because it shows a direct link 
between geometry and the universal aspects of a critical process.
Nevertheless, there still are a few interesting questions that 
deserve a deeper investigation.
For instance, what is the behavior of the non-TT and the scalar 
field perturbation sectors during the splitting of the thick 
brane? 
How does the critical behavior found here change if we consider 
de Sitter or anti-de Sitter branes instead of Poincar\'e branes? 
Is, then, multilocalization of the zero mode possible? 
Which is the effect of a bulk scalar field? 

The author is deeply in debt to C.~F.~Sopuerta 
%for long discussions on different aspects of brane scenarios 
and very thankful to his former collaborators 
K.~Holland and U.-J.~Wiese.
This work has been supported by the Marie Curie Fellowship 
HPMF-CT-1999-00158.

%The author is deeply in debt to C.~F.~Sopuerta for very 
%illuminating discussions on geometric and dynamical aspects 
%of brane world scenarios and also very thankful to his former 
%collaborators K.~Holland and U.-J.~Wiese. 
%This work has been supported by the EU 
%through the Marie Curie Fellowship HPMF-CT-1999-00158.

%\bibliography

\end{document}